\documentclass[twocolumn]{aastex62}

\received{October 24, 2019}
\revised{December 10, 2019}
\accepted{December 11, 2019}

\shorttitle{Impact of Metallicity on Rotation and Activity}
\shortauthors{Amard \& Matt}

\usepackage{graphicx}
\usepackage{amsmath,amssymb}
\usepackage{txfonts}
\usepackage{xcolor,color, colortbl}
\usepackage{ulem}
\usepackage{txfonts}
\usepackage{natbib}
\bibpunct{(}{)}{;}{a}{}{,}
\newcommand{\eg}{{\it e.g. }}

\newcommand{\LA}[1]{{#1}}

\begin{document}

  \title{The Impact of Metallicity on the Evolution of Rotation and Magnetic Activity of Sun-Like Stars}

\correspondingauthor[0000-0002-0786-7307]{L. Amard}
\email{l.amard@exter.ac.uk, s.p.matt@exeter.ac.uk}

\author{Louis Amard}
\affil{University of Exeter \\
Department of Physics \& Astronomy, Stoker Road,\\
Devon, Exeter, EX4 4QL, UK}

\author{Sean P. Matt}
\affil{University of Exeter \\
Department of Physics \& Astronomy, Stoker Road,\\
Devon, Exeter, EX4 4QL, UK}

\begin{abstract}
The rotation rates and magnetic activity of Sun-like and low-mass ($\la 1.4 M_\odot$) main-sequence stars are known to decline with time, and there now exist several models for the evolution of rotation and activity.  However, the role that chemical composition plays during stellar spin-down has not yet been explored.  In this work, we use a structural evolution code to compute the rotational evolution of stars with three different masses (0.7, 1.0, and 1.3~$M_\odot$) and six different metallicities, ranging from [Fe/H]~$=-1.0$ to [Fe/H]~$=+0.5$.  We also implement three different wind-braking formulations from the literature (two modern and one classical) and compare their predictions for rotational evolution.  The effect that metallicity has on stellar structural properties, and in particular the convective turnover timescale, leads the two modern wind-braking formulations to predict a strong dependence of the torque on metallicity.  Consequently, they predict that metal rich stars spin-down more effectively at late ages ($\ga 1$ Gyr) than metal poor stars, and the effect is large enough to be detectable with current observing facilities.  For example, the formulations predict that a Sun-like (solar-mass and solar-aged) star with [Fe/H]~$=-0.3$ will have a rotation period of less than 20 days.  Even though old, metal poor stars are predicted to rotate more rapidly at a given age, they have larger Rossby numbers and are thus expected to have lower magnetic activity levels.  Finally, the different wind-braking formulations predict quantitative differences in the metallicity-dependence of stellar rotation, which may be used to test them.
\end{abstract}

   \keywords{Stars --
   			  Stars:evolution -- 
   			  Stars:low-mass --    			  
                Stars:metallicity --
                Stars:rotation
               }

\section{Introduction}
\label{sect:intro}

Characterizing the evolution of rotation and magnetic activity in low-mass ($\la 1.3 M_\odot$) stars is important for our understanding of stellar evolution in general. 
It allows us to develop methods to determine stellar ages \citep{Barnes2003,MamajekHillenbrand2008, Barnes2010, Angus2015}, to understand stellar variability, and to infer the evolution of the environmental properties (magnetism, winds, and high-energy radiation) of stars.  
It has been known for a long time that the evolution of stellar rotation is a result of changes in their moments of inertia and angular momentum loss by magnetized stellar winds \citep{Schatzman1962, WD1967, Mestel1968, Skumanich72, Mestel1984, Kawaler88, MacGregor2000}.  
Stellar wind properties and magnetism can be probed observationally by chromospheric emission \citep{Noyesetal84, BoroSaikia2018, Egeland2016}, X-rays \citep{Pizzolato2003,Wright2011,Wright2018}, UV \citep{Stelzer2016}, optical variability due to spots \citep{McQuillan2014,BasriNguyen2018,Lanzafame2018} and other magnetic phenomena \citep[Such as white light flares, see][]{Kowalski2009,Stelzer2016}, Zeeman broadening observations of surface flux \citep{Reiners2013}, magnetic field characterisation by Zeeman-Doppler Imaging \citep{Vidotto2016, Reville2016, See2017}, and mass-loss rates \citep{Wood2005}.  
All of these phenomena, generically described as ``magnetic activity'', appear to depend upon stellar structure and age. Most are known to scale with stellar Rossby number \citep{Noyesetal84}, defined as the stellar rotation period, divided by the convective turnover timescale.  
The ultimate origin of magnetic activity in solar-like and low-mass stars appears to be a coupling between the convective envelope \citep{Kippenhahn2012} and the rotation of the star, which produces a dynamo mechanism that generates a magnetic field \citep[see the review by][]{BrunBrowning2017}.

In recent years, many descriptions of angular momentum loss due to magnetized stellar wind have been proposed to explain the observed distribution of low-mass stars' rotation periods  \citep[e.g.,][]{Matt2012, RM12, VSP2013, Brown2014, Mattetal2015, Vansaders2016, Gondoin2017, Garraffo2018, FinleyMatt2018}.  
These models typically include different assumptions for how magnetic fields and mass loss rates vary with stellar properties. 
All these torques are mainly calibrated on open clusters and solar properties, usually assumed to be of the same chemical composition. 
Consequently, none of these description explicitly depends on stellar chemical abundances or metallicity.

Open clusters are indeed very convenient and often used as calibrator, and they have been shown to have a small internal composition dispersion which allows us to treat each of them with a unique chemical mix \citep{Bovy2016}.
For most of them, their mean metallicity is also close to solar.
However, even among the most commonly used clusters, the metallicity can be up to two to three time higher or lower than solar \citep{Netopil2016}. 
The effect of these metallicity differences on the rotation rates has not yet explored for young clusters.

While the study of rotation was initially mostly concentrated on nearby open-cluster members \cite{Skumanich72}, large surveys have opened a new era on studying the rotation and activity of field stars. 
From space, Kepler provided us with 34,000 photometric rotation periods \citep{McQuillan2014}, GAIA DR2 provided 150,000 \citep{Lanzafame2018}, TESS is expected to bring around two hundred thousand \citep{Stassun2019}, and a million more are expected from GAIA DR3 \citep{Lanzafame2018}.  
Ground-based surveys, such as MEarth \citep{Irwin2011,Newton2018}, have also supplied significant numbers of rotation periods.  
Unlike stars in clusters, field-star samples cover a wide range of metallicities and ages, and we often have little information on either of those properties.
However, recent and ongoing spectroscopic surveys such as LAMOST \citep{Cui2012}, the GAIA-ESO survey \citep{Gilmore2012}, GALAH \citep{DeSilva2015} or APOGEE \citep{Majewski2017} are providing us with surface abundances for many of the stars for which we have rotation periods. 
More are coming in the next few years, for example, from the 4MOST \citep{DeJong2019}, MOONS \citep{Cirasuolo2014}, and WEAVE \citep{Dalton2014} instruments.  
Cross-matching these datasets should give a more complete picture, providing accurate masses, rotation periods, metallicities, and, in the best cases, ages.
Therefore, the time is ripe for exploring whether metallicity correlates with rotation or magnetic activity.  
In order to help interpret these data, it is important to develop models that predict whether and how metallicity affects stellar spin-down.

\cite{Cortes2009} looked for correlations between metallicity and observed $v\sin{i}$ values. They did not find any significant trends, but their sample included field stars at very mixed evolutionary stages from the main sequence to the horizontal branch. \LA{More recently, \cite{LorenzoOliveira2016} derived an empirical age-mass-metallicity-activity relation and tested it on an old open cluster, obtaining reasonably good results. They advertise that color is not a good enough tracer of the age-activity relation and both mass and metallicity need to be accounted for.} 
To date, there has been little theoretical work exploring the effects of metallicity on low-mass star rotation.
\cite{Ekstrom2008} studied the effect of metallicity on massive stars evolution, including rotation. 
However, the structure of low-mass stars is very different, and the processes governing their rotational evolution are not the same. 
The low-mass-star models of \cite{Vansaders2016,Vansaders2019} do account for different metallicities, but they did not systematically study the effects. 
\cite{Amard2019} presented a new grid of low-mass stellar evolution models, which included rotation and a range of metallicities.
They found that metallicity had a strong affect on rotation, modifying the spin-up during the pre-main-sequence contraction phase, as well as the spin-down on the main sequence.
However, they did not provide an in-depth analysis, nor did they identify the key factors responsible for modifying the spin-down.  
They also computed the evolution using only one particular formulation for the magnetic stellar wind torque.

In this paper, we aim to characterize, quantify, and understand the effects of metallicity on the rotational evolution, as well as to explore the predictions of different stellar-wind braking formulations from the literature.
We use an evolution code that includes a fully-consistent treatment of rotation, to test three angular momentum loss descriptions at different metallicities, from [Fe/H]~$=-1$ to [Fe/H]~$=+0.5$.  
We also compute the evolution of stars with three different masses, 0.7, 1.0, and 1.3~$M_\odot$, in order to disentangle the effects of mass and metallicity.
Section \ref{sect:model} of the paper is is dedicated to the description of the rotational evolution models, including the three different magnetised wind torques we explore.  
In \S~\ref{sect:struc}, we show how metallicity affects the stellar properties that are important for the wind torque and rotational evolution.
In \S~\ref{sect:evolrot}, we present the effect of metallicity on the rotational evolution for different masses and torque descriptions. We discuss the results in \S~\ref{sect:discussion} and summarize and conclude in \S~\ref{sect:conclusion}.


\section{Description of the models}
\label{sect:model}
\subsection{Evolution models (and chemical compositions at different Z)}

  We use the STAREVOL code to compute the evolution of stars. Its last version has been shown to be in good agreements with other standard evolution codes and fully operational to compute in a self-consistent way the rotational evolution at the same time as the structural evolution \citep[see][for more details]{Siess2000,Palacios2003,PalaciosCharbonnel2006,DecressinMathis2009,Lagarde2012,Charbonnel2013,Amard2016,Amard2019}.
  We compute models of stars with 0.7, 1.0, and 1.3 M$_\odot$, at six metallicities each, and with three braking laws, as described below.  The metallicities and the corresponding Helium abundances are presented in table~\ref{tab:abund}. The $\alpha$-enrichment at low-metallicity reflects the composition observed in the thin disc of our galaxy \citep[see \eg][]{Furhmann2011}.  We assume the stars rotate as solid bodies and account for the centrifugal effect on the hydrostatic potential of the star following \cite{ES76} formalism.

  Our models all start with the same initial conditions. This is for simplicity and because we have too little constraint on how the initial spin rates might vary (or not) with metallicity.  
  We start the models at a fully convective stage on the Hayashi branch with a period of 2.3 days and a ``disc-locking'' time of 4 Myr, during which the period is assumed to be constant (but allowed to vary thereafter). 
  This set of parameter leads to some relatively fast rotators ($\approx$100 km.s$^{-1}$ at the ZAMS for the M$_\odot$, Z$_\odot$ model), which allows us to study the transition from the saturated (rapidly rotating) to unsaturated (slowly rotating) regime. 
  The choice of initial conditions becomes unimportant after an age of $\sim$~1~Gyr, as stars of all initial spin rates converge onto a common sequence \citep{BouvierPPVI}.  
  We stop the evolution of the models at the end of the main sequence, when the central hydrogen abundance reaches 10$^{-7}$ its initial value. 
  We do not model the post-main-sequence phase, which is characterized by a radius expansion that dominates the rotational evolution.

  \begin{table}[h]
    \begin{center}
    \caption{Chemical abundances and metallicities adopted in the present study, scaled according to the solar chemical mixture by \citet{AsplundGrevesse2009}}
    \begin{tabular}{ c c c c c }
      \hline
      \hline
      [Fe/H] &  [$\alpha$/Fe] &    $Z$     & $Y$ \\
      \hline
      +0.5 & 0.0 & 0.03866  & 0.3090 \\
      +0.3 & 0.0 & 0.02565  & 0.2884 \\
      0.0  & 0.0 & 0.013446 & 0.2691 \\
      -0.3 &+0.1& 0.00796   & 0.2577 \\
      -0.5 &+0.2& 0.00593   & 0.2533 \\
      -1.0 &+0.3& 0.00224   & 0.2493 \\
      \hline
    \end{tabular}
    \label{tab:abund}
    \end{center}
  \end{table}

  \subsection{Braking laws}
  \label{sect:torques}

    We will primarily use and compare two modern braking laws, \citet[][hereafter vSP]{VSP2013} and \citet[][hereafter M15]{Mattetal2015}. Both of these have been shown to reproduce some of the broad structures seen in observed distributions of rotation periods versus mass (or color) in young clusters and field stars \citep{Mattetal2015, Amard2016, Vansaders2016, Cellier2016, Coker2016, Agueros2018, Douglasetal2017, Vansaders2019, Curtis2019}.
    Both of these braking laws were derived from the stellar wind simulations of \cite{Matt2012}, but each of them making slightly different assumptions about how the mass loss rates and surface magnetic field strengths scale with stellar properties. \citet[][hereafter K88]{Kawaler88} is a classical braking law that has been used (in various forms) in several previous works, and we include it for comparison.  Note that none of these braking laws were derived in particular to anticipate the effects of metallicity, but we will show how each formulation does imply a dependence of the rotational evolution on metallicity.

    The three torques each have a constant factor that we have calibrated to reproduce the solar rotation rate, assumed to be $\Omega_\odot/2\pi=0.455 \, \mu$Hz, to a tolerance of \LA{0.03} $\Omega_\odot$, for the models with solar metallicity and 1.0 $M_\odot$. 
    All three torques also have a bifurcated form, with one formulation for the magnetically ``saturated'' regime and one for the ``unsaturated'' regime.  
    The location of the transition from saturated to unsaturated forms is approximately fit to best reproduce the rapidly rotating solar mass stars in young open clusters, and we adopt the same value (at solar metallicity for 1.0 $M_\odot$) for each formulation.
    
    \LA{The K88 torque assumes the saturated/unsaturated transition occurs at a fixed angular rotation rate.  By contrast, the two modern braking laws (vSP and M15) assume that the transition occurs at a fixed Rossby number for all stars.  This Rossby scaling is justified both by the fitting of observed rotation rates in young clusters \citep{Krishnamurthi97} and on an assumed relationship between torques and magnetic activity, which do seem to show a saturated/unsaturated transition at constant Rossby number \citep{Noyesetal84, Pizzolato2003, Wright2011, Wright2018}.  However, it is possible that the saturation of the torque depends on properties other than the Rossby number, which we neglect here.}

    

  \subsubsection{\cite{Mattetal2015}}

    The \cite{Mattetal2015} description of angular momentum extraction is based on the formalism of \cite{Matt2012}, with the assumptions that 1) the mass loss rate and the magnetic field strength both depends only on the stellar Rossby number, 2) the torque has an arbitrary extra mass-dependency to better reproduce the low-mass end of the observed rotational evolution. It is given by
    \begin{equation}
    \frac{\mathrm{d}J}{\mathrm{d}t} = -K_\textrm{M} \left(\frac{R_\star}{R_\odot}\right)^{3.1}\left(\frac{M_\star}{M_\odot}\right)^{0.5}\beta \frac{\Omega_\star}{\Omega_\odot}
    \left(\frac{\tau_{cz}\Omega_\star}{\tau_{cz \odot}\Omega_\odot}\right)^2 \rightarrow {\rm unsaturated},
    \label{eq:m15_unsat}
    \end{equation}
    \begin{equation}
    \frac{\mathrm{d}J}{\mathrm{d}t} = -K_\textrm{M} \left(\frac{R_\star}{R_\odot}\right)^{3.1}\left(\frac{M_\star}{M_\odot}\right)^{0.5}\beta \frac{\Omega_\star}{\Omega_\odot}
    \left(\frac{Ro_\odot}{Ro_{\rm sat}}\right)^2  \rightarrow  {\rm saturated},
    \label{eq:m15_sat}
    \end{equation}
    where $R_\star$, $M_\star$, and $\Omega_\star$ are the stellar radius, mass, and (solid-body) angular rotation rate.  All values with the subscript ``$\odot$'' are values for the present-day sun.  The factor $K_\textrm{M} = 6.7 \times 10^{30}$ erg is the calibration constant.
    
    This formalism depends on the Rossby number in the convective envelope, defined as
    \begin{eqnarray}
    Ro = P_{\rm rot,\star}/\tau_{\rm cz}, \\
    \frac{Ro}{Ro_\odot} =
    \frac{P_{\rm rot,\star} \tau_{\rm cz \odot}}{\tau_{\rm cz} P_{\rm rot,\odot}} =
    \frac{\tau_{\rm cz \odot}\Omega_\odot}{\tau_{\rm cz}\Omega_\star}
    \label{Eq:Rossby}
    \end{eqnarray}
    with $P_{\rm rot,\star}$ the mean surface rotation period ($\equiv 2\pi/\Omega_\star$), and $\tau_{cz}$ the turnover timescale, defined by the local pressure scale-height divided by the local convective velocity, computed at a position of half a scale height above the base of the convection zone.  In this model, the magnetic field saturates when $Ro < Ro_\textrm{sat}$, and this saturation value is determined by imposing that our 1M$_\odot$ roughly reproduces the dispersion in rotation periods in ZAMS clusters.  This requires $\chi = \frac{Ro_\odot}{Ro_{\rm sat}} = 11$, which we adopt throughout\footnote{We adopt a slightly different value of $\chi$ than M15 (they used $\chi=10$) because \citet{Amard2019} found it was a better fit to data.  We use the same saturation value for all 3 torque prescriptions.}.

    The factor
    \begin{equation}
      \label{eq:beta}
      \beta \equiv \left[1 + 193 \left(\frac{\Omega_\star^2     R_\star^3}{GM_\star}\right)\right]^{-m}.
    \end{equation}
    models the effect of magneto-centrifugal acceleration of the wind, and it is only important (significantly different from unity) for very rapid rotators\footnote{M15 did not include the factor $\beta$, for simplicity, but our formulation follows from \cite{Matt2012}.  We include the $\beta$-factor here for completeness.}.  \cite{Matt2012} found $m=0.222$, which we adopt here. 

  \subsubsection{\cite{VSP2013}}

    The vSP torque is also based on the formalism of \cite{Matt2012}, but with the assumptions that (1) the mass loss rate scales with the stellar luminosity, $L_\star$, (2) the magnetic field strength scales with the square root of the pressure at the stellar photosphere \LA{taken at $\tau = 2/3$ from the stellar structure computation}, $P_\textrm{eff}$, and (3) both have a power-law scaling with Rossby number \LA{in the unsaturated regime such that they obtain the same scaling as in \cite{Matt2012}}.
    The torque is given by\footnote{\cite{VSP2013} set $\beta=1$, for simplicity (they called it $c(\omega)$), but we include the full formulation here.}
    \begin{equation}
      \label{eq:vsp_unsat}
      \frac{\mathrm{d}J}{\mathrm{d}t} = -K_\textrm{vSP} \left(\frac{R_\star}{R_\odot}\right)^{3.1} \left(\frac{M_\star}{M_\odot}\right)^{-0.22} \lambda \ \beta \ \frac{\Omega_\star}{\Omega_\odot}
      \left(\frac{\tau_{\rm cz}\Omega_\star}{\tau_{\rm cz \odot}\Omega_\odot}\right)^2  \rightarrow {\rm unsaturated},
    \end{equation}
    \begin{equation}
      \label{eq:vsp_sat}
      \frac{\mathrm{d}J}{\mathrm{d}t} = -K_\textrm{vSP} \left(\frac{R_\star}{R_\odot}\right)^{3.1} \left(\frac{M_\star}{M_\odot}\right)^{-0.22} \lambda \ \beta \ \frac{\Omega_\star}{\Omega_\odot} \left(\frac{Ro_\odot}{Ro_{\rm sat}}\right)^2  \rightarrow  {\rm saturated},
    \end{equation}
    with
    \begin{equation}
      \label{eq:lambda}
      \lambda = \left(\frac{L_\star}{L_\odot}\right)^{0.56} \left(\frac{P_\textrm{eff}}{P_{\odot}}\right)^{0.44}.
    \end{equation}
    The calibration constant $K_\textrm{vSP} = 7.5 \times 10^{30}$ erg.  The only differences between the vSP torque formulation and that of M15 is in the calibration constant (chosen in both cases to match the sun), a different power on the factor of $M_\star$, and the vSP torque uniquely includes the $\lambda$-factor\footnote{\citet{Vansaders2016} suggested a modification to the vSP torque by adding a second Rossby threshold, above which angular momentum loss becomes null (or negligible).  For the sake of comparison between torque models, we do not include this effect here.}.

  \subsubsection{\cite{Kawaler88}}

    We also compute models using the torque of Kawaler and have included a saturation angular velocity following \citet{Chaboyeretal1995a}.
    \begin{equation}
      \label{eq:k88_unsat}    
      \frac{\mathrm{d}J}{\mathrm{d}t} = -K_\textrm{K} \left(\frac{R_\star}{R_\odot}\right)^{1/2}\left(\frac{M_\star}{M_\odot}\right)^{-1/2}\left(\frac{\Omega_\star}{\Omega_\odot}\right)^3   \rightarrow  {\rm unsaturated},
    \end{equation}
    \begin{equation}
      \label{eq:k88_sat}    
      \frac{\mathrm{d}J}{\mathrm{d}t} = -K_\textrm{K} \left(\frac{R_\star}{R_\odot}\right)^{1/2}\left(\frac{M_\star}{M_\odot}\right)^{-1/2} \frac{\Omega_\star}{\Omega_\odot}\left(\frac{\Omega_\textrm{sat}}{\Omega_\odot}\right)^2 \rightarrow  {\rm saturated},
    \end{equation}
    Where $K_\textrm{K}=6.3 \times 10^{30}$ erg 
    is calibrated to reproduce the solar case and $\Omega_\textrm{sat} = 11 \Omega_\odot$.  Note that this torque has a substantially weaker dependence on stellar properties, and the transition between the saturated and unsaturated regimes occurs at a constant rotation rate for all models (as opposed to having a Rossby scaling in the other two torques).


\section{How Metallicity Affects Structure and Braking Laws}
\label{sect:struc}

  \begin{figure}
    \centering
    \includegraphics[height=7.2cm]{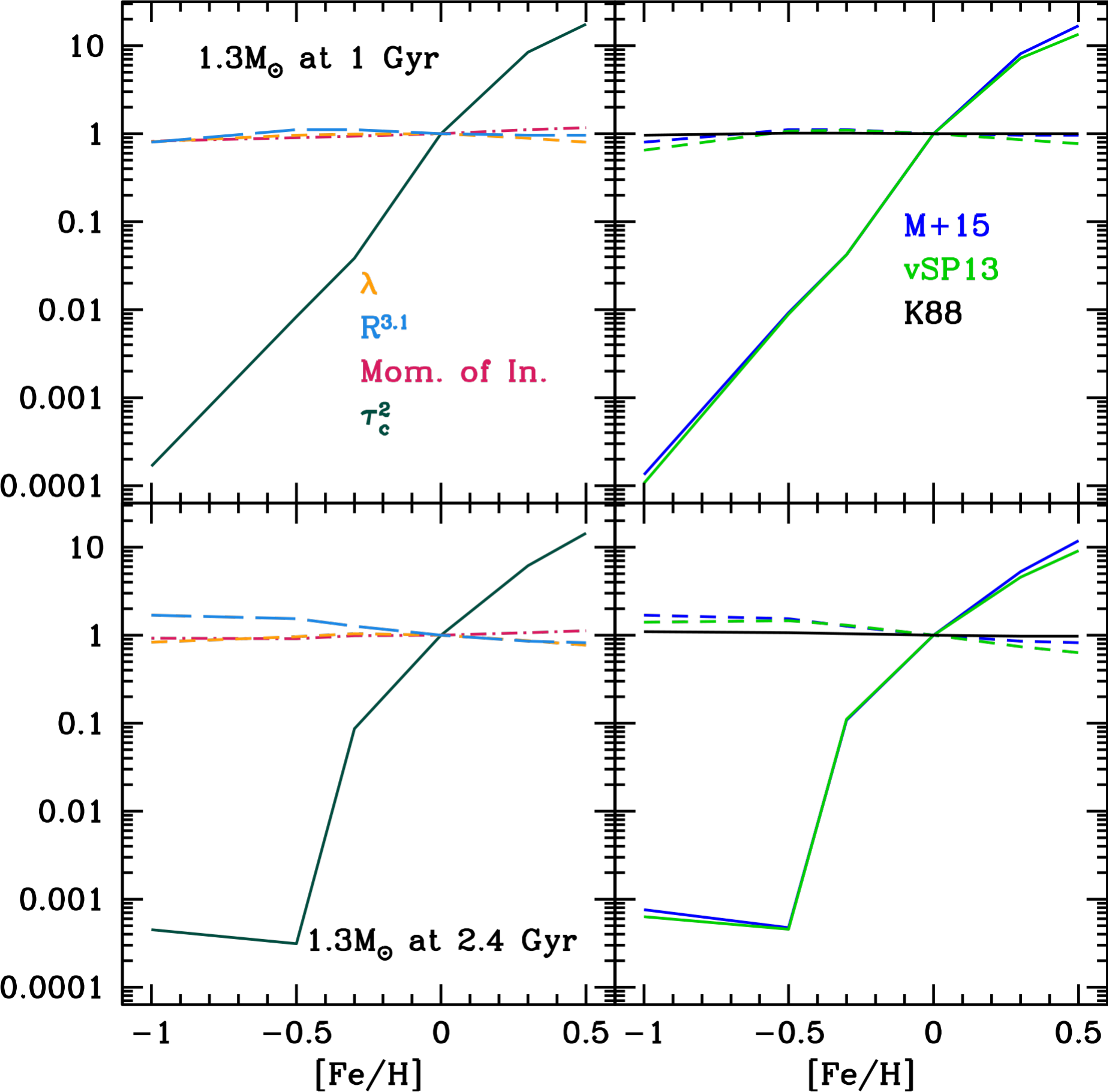}
    \vspace{0.1cm}
    \includegraphics[height=7.2cm]{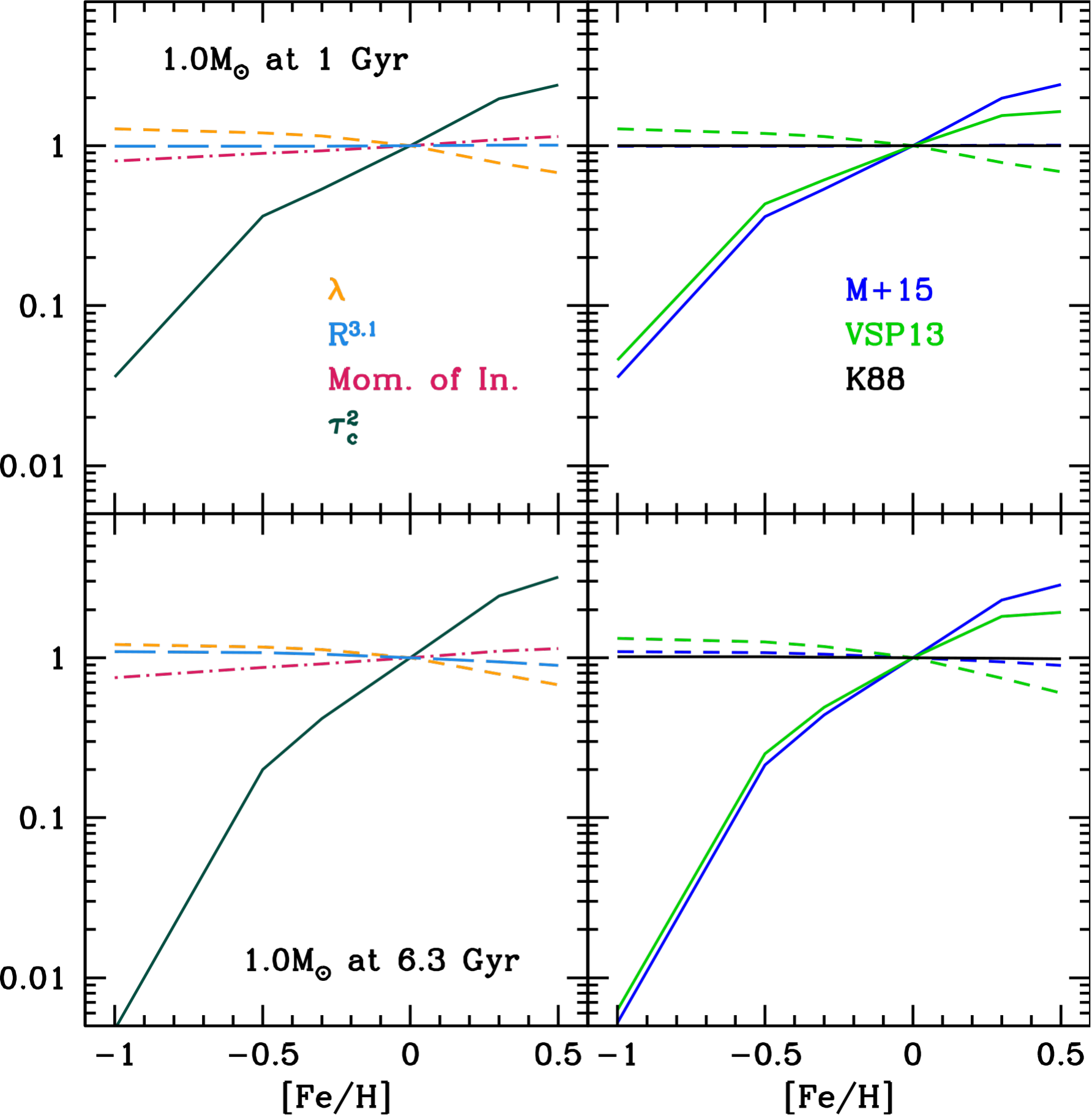}
    \vspace{0.1cm}
    \includegraphics[height=7.2cm]{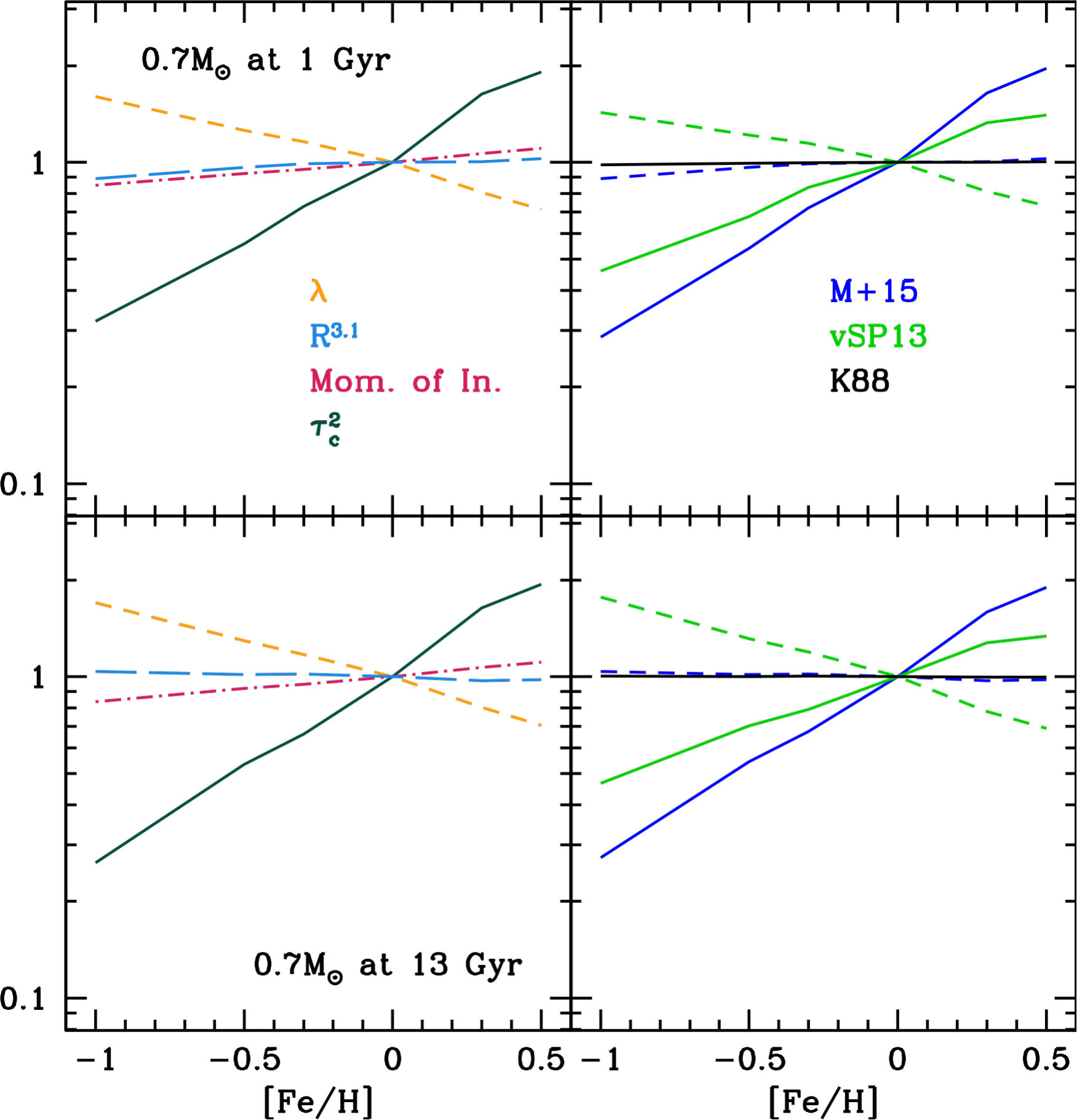}
    \caption{Left column: Stellar structural quantities (as indicated) that are important for spin-down, as a function of metallicity.  Right column: Saturated (dashed lines) and unsaturated (solid lines) torques for all three braking laws considered (see \S \ref{sect:torques}), as a function of metallicity.  All quantities are normalized to their value at solar metallicity.  The top, middle, and bottom sets of four panels are for models with 0.7, 1.0 and 1.3$M_\odot$.  For each mass, the top and bottom rows show two different ages.  See the text of \S \ref{sect:struc} for a full description and discussion.} \label{Fig:param}
  \end{figure}

  None of the braking laws currently in use contain an explicit dependence on metallicity.  
  However, the braking laws depend on various stellar structural properties (as described in \S \ref{sect:torques}) that are influenced by metallicity.  
  It is well known that a higher abundance of elements heavier than helium increases the global opacity of the star and so leads to an increase of the temperature gradient in the star. This leads to a deeper convective envelope (e.g., as can be seen in fig.\ 11 of \citealp{Amard2019}), slower convective speeds, and consequently a longer convective turnover timescale \citep{KitchatinovOlemskoy2011,Charbonnel2017,Karoff2018}.  
  Also, a higher abundance of metals results in a cooler, dimmer star, with a lower central temperature, which results in a longer main-sequence lifetime.

  Figure~\ref{Fig:param} presents the variation of the most relevant stellar properties (left panels) and torques (right panels) with metallicity, for all three stellar masses considered (top, middle, and bottom sets of 4 panels are for 0.7, 1.0, and 1.3 $M_\odot$, respectively).  
  For each mass, the two top panels show the quantities at 1 Gyr, and the two bottom panels show the same quantities at an age corresponding to when the most metal poor model has its central hydrogen abundance below 10$^{-5}$ for the 1.3 and 1.0 $M_\odot$ stars and below 0.2 for the 0.7$M_\odot$ star (at an age of 13, 6.3, and 2.4 Gyr, for 0.7, 1.0, and 1.3 $M_\odot$).  To focus on variations with metallicity, the quantities shown in each panel are normalized to their value at the solar metallicity ($[Fe/H]=0$).

  The \LA{dot-dash maroon} line in the left panels of Figure~\ref{Fig:param} show the moment of inertia of the star, which determines how quickly external torques can change the star's rotation rate.  The moment of inertia varies with the stellar mass and radius (squared), but also with the gyration radius \citep{Rucinski1988} and thus on the mass distribution in the star.  For all masses, the trend is that metal poor stars have a smaller radius and gyration radius at a given evolutionary point.  Nevertheless, this variation with metallicity at a given age is relatively small (at most a few tens of percent variation over the metallicity range considered), and so the effect on rotational evolution will be similarly small.
 
  The \LA{long-dashed blue} line in the left panels of Figure~\ref{Fig:param} shows $R_\star^{3.1}$.  For a given mass and rotation rate, this is the most important factor in the saturated-regime torque of M15 (eq.\ [\ref{eq:m15_sat}]) that depends on metallicity.  The dashed blue line in the right panels shows the M15 saturated torque, which follows the same behavior as $R_\star^{3.1}$.  It is clear from the figure that the radius component (even when raised to the exponent $3.1$) and the saturated-M15 torque don't vary much with metallicity at a given age. During most of the main sequence phase, the lower the Z, the smaller the radius. However, stellar radius increases during main sequence evolution, and the lower metallicity stars evolve more rapidly.  Thus, in the bottom panel for each mass, the Figure shows an increase in stellar radius toward the lowest metallicities, as those stars approach the subgiant branch.  
  In the case of the K88 torque, shown as the black line in the right panels of Figure~\ref{Fig:param}, both the saturated- and unsaturated-regime torques (eqs.\ [\ref{eq:k88_unsat}] and [\ref{eq:k88_sat}]) vary with $R_\star^{0.5}$, giving an extremely weak dependence on metallicity for all masses and at all times.

  The \LA{solid dark} green line in the left panels of Figure~\ref{Fig:param} shows $\tau_{\rm cz}^2$, which is a factor that appears in the unsaturated-regime for both the M15 and vSP torques (eqs.\ [\ref{eq:m15_unsat}] and [\ref{eq:vsp_unsat}]). 
  It is clear that the turnover timescale is very sensitive to metallicity and mass, with $\tau_{\rm cz}^2$ spanning a factor of $\sim$7 for the 0.7 $M_\odot$ star and $\sim$5 orders of magnitude for the 1.3 $M_\odot$ star, over the metallicity range considered.  Furthermore, $\tau_{\rm cz}^2$ changes substantially toward the end of the main-sequence phase, which is most clear for the 1.0 and 1.3 $M_\odot$ models at the lowest metallicities.  \LA{The $\tau_{\rm cz}^2$-factor dominates the metallicity-dependence of the unsaturated torques in both the M15 and vSP formulations.  Recall that the dependence of these torques on $\tau_{\rm cz}^2$ is due to an implicit assumption of how convection affects the dynamo action (and thus the surface magnetic fields and mass loss rates).} The unsaturated-regime torque of M15 (eq.\ [\ref{eq:m15_unsat}]) contains a factor of $\tau_{\rm cz}^2 R_\star^{3.1}$ and is shown as a solid blue line in the right panels of the Figure. 
  By contrast with the saturated-M15 torque, the unsaturated-M15 torque has a strong dependence on metallicity, where metal rich stars are predicted to undergo a stronger stellar wind torque than metal poor stars.

  For a given mass and rotation rate, the only difference between the M15 and vSP torques is that the latter contains a factor of $\lambda= L_{\star}^{0.56} P_{\star}^{0.44} $, which is shown with \LA{a yellow} line in the left panels of Figure~\ref{Fig:param}.  
  Both the luminosity and the photospheric pressure decrease with increasing metallicity because the trend is for stars to become less compact and cooler.  
  However, more metal-poor and massive stars have a higher radiative pressure, which decreases the effective pressure and explains the flattening of lambda at low metallicity, in the 1.3M$_\odot$ diagram. Thus the largest contrast appears for low mass models, where $\lambda$ ranges over a factor of $\sim$2.  
  As a consequence of the $\lambda$-dependence, the unsaturated-vSP torque (solid green line in right panels) depends on metallicity in the same way as that of M15, but with a shallower dependence.  Furthermore, in the saturated regime, the vSP torque has a weak inverse dependence on metallicity.

  In summary, the moment of inertia has only a small dependence on metallicity, so we expect the largest change in rotational evolution will be due to the difference in external torques.  The K88 torque has almost no dependence on metallicity.  The unsaturated-regime torques of M15 and vSP, which are dominated by a factor of $\tau_{\rm cz}^2$, depend similarly on metallicity, with the vSP having a slightly shallower dependence.  Finally, while saturated-regime torque of M15 has very little dependence on metallicity, that of vSP has a weak dependence that is in the opposite sense as in the unsaturated regime.


\section{Evolution of Rotation Rates}
\label{sect:evolrot}

  In this section, we describe the rotation evolution of the modeled stars.  The focus here is on the phenomenology of how metallicity and adopted braking law affect the predicted rotational evolution, which is a consequence of the trends discussed in section \ref{sect:struc}.  Figures \ref{fig:Evol_1Msun} and \ref{fig:Omega_vs_t_all} show the evolution of rotation rate for all models, which we describe and discuss in the following subsections.

  \subsection{Solar Mass, Late-Stage Evolution}
  \label{sect:evolrot_late}

    \begin{figure}
      \includegraphics[width=0.48\textwidth]{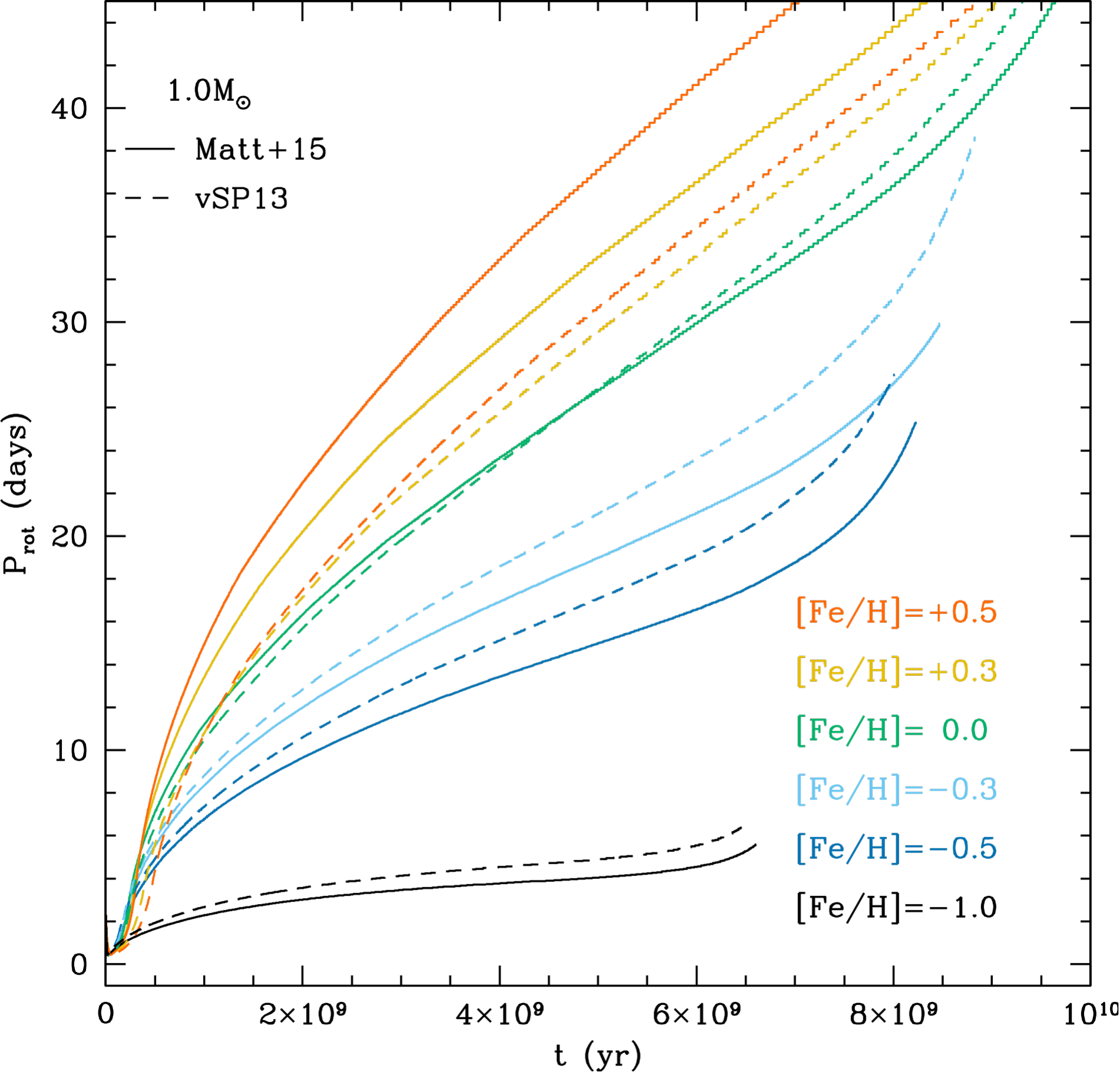}
      \caption{Rotation period as a function of time for a star with  1.0 solar mass.  Models are shown for metallicities of [Fe/H]=-1.0, -0.5, -0.3, 0.0, +0.3 and +0.5 (from bottom to top and color coded as indicated).  The solar case is shown in green.  Solid lines show models that use the stellar-wind-torque formulation of \cite{Mattetal2015}, and dashed lines use that of \cite{VSP2013}. At ages $\ga 1$~Gyr, stars with higher metallicity are predicted to spin down more effectively, and the magnitude of this affect depends upon the torque formulation.}
      \label{fig:Evol_1Msun}
    \end{figure}

    Figure~\ref{fig:Evol_1Msun} shows the time-evolution of the rotation period of a one solar-mass model in linear-linear space, which emphasizes the main sequence evolution beyond an age of $\sim$1~Gyr, during which stars are in their unsaturated phase of spin-down.  The solid and dashed lines show the prediction for the braking laws of M15 and vSP, respectively, and all metallicities are shown with different colors (as indicated).  The green lines show the evolution for a star with solar metallicity.  Both braking models show an approximately Skumanich-type spin-down \citet[][$P \propto t^{1/2}$]{Skumanich72}, and both models predict the approximate value of solar rotation at the solar age ($\approx 4.57$ Gyr), because both have been designed and calibrated to do so.

    Both braking models predict that stars with higher metallicities spin down much more efficiently than those with lower metallicity. The effect is remarkably strong for both torques, particularly toward low-metallicity cases.  The trend with metallicity is predicted to be somewhat weaker for the vSP torque, as compared to the M15 torque, due to the differences in their torque formulations (see \S \ref{sect:struc}).  For a quantitative comparison, Table~\ref{tab:Prot_sun} lists the predicted rotation period at the age the Sun, for all metallicities and all three angular momentum loss models.
   
    Finally, note that the model tracks terminate at different times because the duration of the main-sequence phase is longer for stars with higher metallicity.  Thus, metal poorer stars are predicted to spin more rapidly at the end of the main-sequence phase because of both a reduced spin-down rate and a reduced duration of spin-down.

    \begin{table}
      \caption{Rotation period of the 1 M$_\odot$ models at the age of the Sun (4.57 Gyr) for different metallicities.}
      \label{tab:Prot_sun}
      \centering
      \begin{tabular}{c c c c}
      \hline\hline
      Metallicity & vSP13 & M+15 & K88 \\
      $\left[\mathrm{Fe/H}\right]$ & (days) & (days) & (days) \\
      \hline
      -1.0 & 4.5 & 3.9 & 27.8 \\
      -0.5 & 16.1 & 14.4 & 25.9 \\
      -0.3 & 19.7 & 18.1 & 25.5 \\
      0.0 & \LA{25.5} & \LA{25.6} & 24.7 \\
      0.3 & 27.5 & 31.4 & 23.1 \\
      0.5 & 28.6 & 35.4 & 22.4 \\
      \hline
      \end{tabular}
    \end{table}

    \subsection{Solar Mass, Early Stages}

    \begin{figure*}
      \centering
      \includegraphics[width=\textwidth]{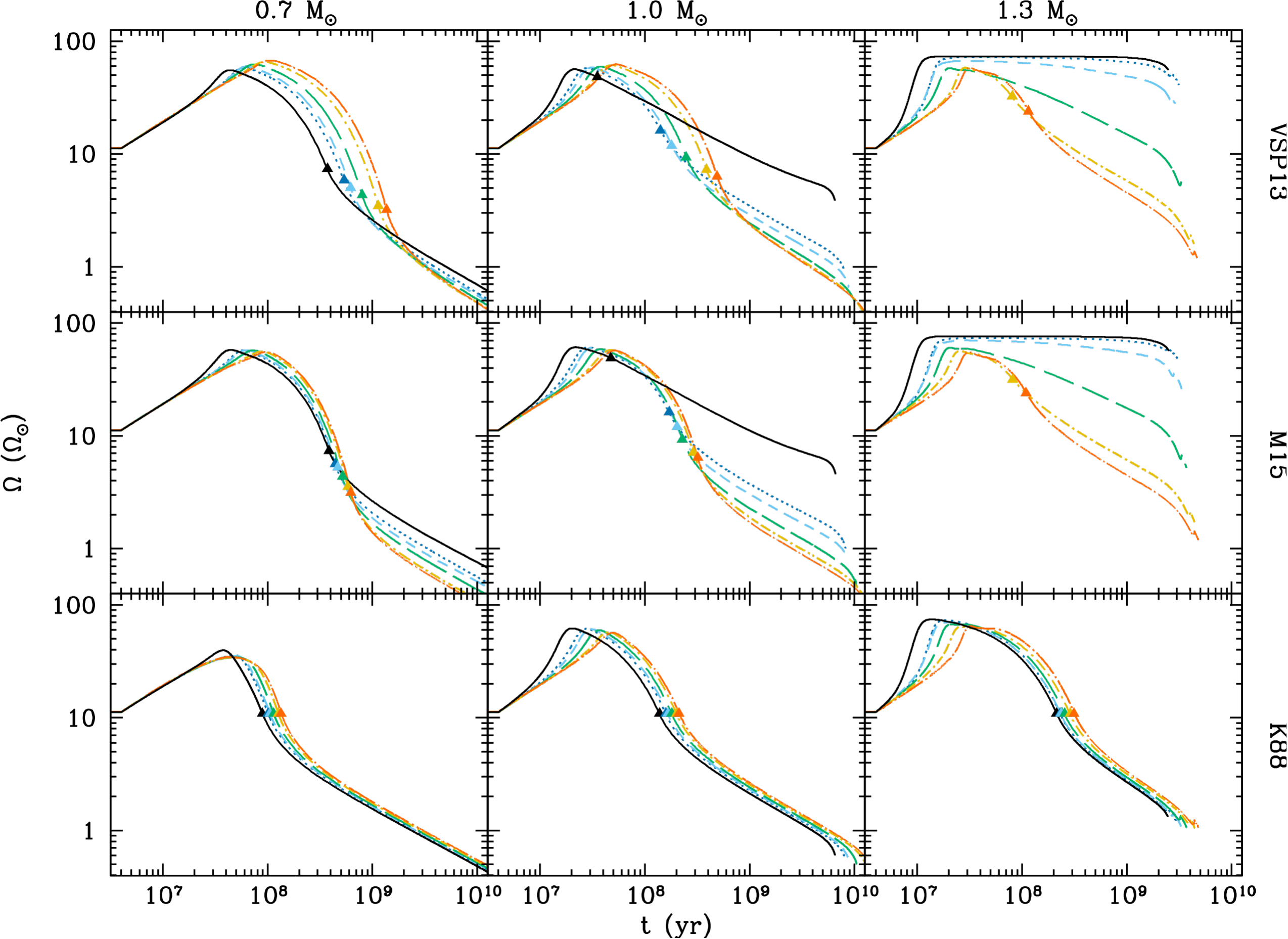}
      \caption{Evolution of the surface angular velocity as a function of time for all models. In each panel, six metallicities are shown, [Fe/H]=-1.0, -0.5, -0.3, 0.0, +0.3 and +0.5, respectively \LA{in black solid, blue dots, light-blue short dashed, green long dashed, yellow short-dashed-dot, and red long-dashed-dot lines (see colour code of Fig.~\ref{fig:Evol_1Msun}).}  The left, middle, and right columns show models for stars with 0.7, 1.0, and 1.3 solar masses, respectively.  The top, middle, and bottom rows show models using the torque of \cite{VSP2013}, \cite{Mattetal2015}, and \cite{Kawaler88}, respectively. \LA{The triangles indicate for each model the transition from the saturated to unsaturated regime when it exists.} The predicted effect of metallicity on rotation is complex, depending on the evolutionary phase and the adopted torque prescription.}
      \label{fig:Omega_vs_t_all}
    \end{figure*}

    Figure~\ref{fig:Omega_vs_t_all} shows the evolution of the angular velocity (using a log-log scale) for the entire range of parameters we covered. The top-middle (for vSP) and middle-middle (for M15) panels of Figure~\ref{fig:Omega_vs_t_all} show the same models as Figure~\ref{fig:Evol_1Msun}, but shown as the angular spin rate (instead of Period) versus time and on a log-log scale. While we still see that the late (unsaturated) stage evolution is marked by an approximately Skumanich spin-down, with the metallicity trends discussed in section~\ref{sect:evolrot_late}, the space in Figure~\ref{fig:Omega_vs_t_all} more clearly shows the complex evolution at earlier stages ($\la 1$ Gyr). 

    Before they reach the ZAMS, the model stars spin up, due to contraction.  Once they arrive on the MS, the structure stabilizes so that the torque acts to decrease the spin rate.  Already during the spin-up phase, the metallicity makes a difference in the evolution. Metal-poor stars contract faster and, for a given set of initial conditions, reach their maximal velocity at a younger age. They then begin their spin-down phase sooner than more metal-rich stars.  As a consequence of the earlier onset of spin-down, metal poor stars are predicted to rotate more slowly than the metal rich stars, during the early (saturated) phase of spin-down (this assumes a fixed initial condition).  When compared to the M15 model (middle-middle panel), the vSP model (top-middle panel) predicts a stronger trend for metal poor stars to rotate more slowly than metal rich stars, in this early phase.  As shown in section~\ref{sect:struc}, this is because the vSP torque has a significant inverse-dependence on metallicity in the saturated regime.

    In this early spin-down phase, all models remain in the saturated-torque regime until they spin down enough to transition to the unsaturated regime. 
    \LA{The saturated-unsaturated transition is noted on Figure~\ref{fig:Omega_vs_t_all} with a triangle, it roughly corresponds to the time at which the spin tracks in Figure~\ref{fig:Omega_vs_t_all} change from a negative to a positive curvature around a few hundred Myr.} The rotation rate at which the transition occurs varies for models using the M15 and vSP torques because the transition occurs at a constant Rossby number.  A metal-enriched star bears a deeper convective envelope with a longer turnover timescale, and thus reaches a given Rossby number at a lower rotation rate.  By contrast, the models using the K88 torque all make the saturated-unsaturated transition at a constant rotation rate.

    The bottom middle panel of Figure~\ref{fig:Omega_vs_t_all} shows the evolution of a solar-mass star predicted by the K88 braking law.  Relative to the other two braking laws, the K88 formulation predicts a much smaller overall dependence on metallicity.  In the late (unsaturated) phase, the metallicity trend is opposite of that predicted for the other two braking laws.  The trend is mainly due to the weak metallicity-dependence of the moment of inertia, discussed in section~\ref{sect:struc}.

    In summary, after the disc-coupling phase and before the end of the main sequence, metallicity appears to affect the rotational evolution in three phases.  First, metal poor stars spin up more rapidly than metal rich stars, simply due to a more rapid contraction phase, and thus reach the spin-down phase quicker.  Second, the transition between the saturated and unsaturated regime is strongly affected by Z. The vSP and M15 braking laws (top-middle and middle-middle panels of Fig.~\ref{fig:Omega_vs_t_all}) predict that during the early (saturated) phase of spin-down, metal poor stars will spin more slowly than metal rich stars (for a fixed initial condition).  Finally, the same two braking laws predict that in the late (unsaturated) spin-down phase, metal poor stars spin down less effectively than metal rich stars, reversing the trend of the saturated phase.

  \subsection{Variations with Stellar Mass}

    The evolution of rotation is strongly affected by the stellar mass \citep[see, \textit{e.g.},][]{Mattetal2015}. To test how metallicity affects the evolution at different masses, we compute models of one solar mass, one lower mass (0.7M$_\odot$) and one higher mass (1.3M$_\odot$).  The model at 1.3M$_\odot$ at solar metallicity is very close to the so-called "Kraft break" \citep{Kraft1967}, beyond which higher-mass stars do not seem to spin-down effectively. A 0.7M$_\odot$ star has a deep convective envelope and a main sequence lifetime longer than the age of the universe (and thus a very slowly varying structure).  The left, middle, and right columns of Figure~\ref{fig:Omega_vs_t_all} shows the models with 0.7, 1.0, and 1.3 $M\odot$, respectively.

    The bottom row in the Figure shows models using the K88 braking law.  It is clear that all the models computed with this torque have a very similar rotational evolution.  The K88 torque depends only weakly on stellar mass (with a factor $R_\star^{0.5}M_\star^{-0.5}$), so the main difference seen in the rotational evolution is due to differences in the pre-main-sequence contraction times and moments of inertia (see \S~\ref{sect:struc}).  Note that K88 is a classical models that does not well fit the observed rotation distributions for stars of non-solar mass, unless those models are tuned for each mass (which we have not done here).  Thus, we only include the predictions from the K88 formulation for completeness and comparison to the other braking laws.
     
    For the vSP and M15 braking laws, the 0.7 $M_\odot$ models in Figure~\ref{fig:Omega_vs_t_all} show an evolution that is qualitatively similar to the solar-mass cases (middle panels).  They both show the same trends with metallicity, at various phases of their evolution.  The main differences in this mass range are due to previously known trends (see, e.g., M15) for (a) lower mass stars to transition to the unsaturated regime at later ages, and (b) lower mass stars to tend toward slower rotation rates than higher mass stars, in the unsaturated regime.

    The 1.3M$_\odot$ models (right column in Fig.~\ref{fig:Omega_vs_t_all}) using the M15 and vSP braking laws (top two panels) show the largest sensitivity to the stellar composition.  While the metal-rich models spin down following a Skumanich-like trend, the three most metal poor models are not significantly braked and remain fast rotators for their whole main-sequence lifetime.  This extreme sensitivity to the metallicity is attributed to the rapid shrinking of the convective envelope (and thus decrease in $\tau_{\rm cz}$) as one considers stars with higher mass or lower metallicity (discussed further in \S~\ref{sect:rossby}).  In other words, metallicity affects the mass at which the Kraft break occurs, and our most massive and lowest metallicity models are near or beyond the Kraft break.  Indeed, for [Fe/H]=-1.0, even the solar-mass model (black line in middle column) appears to be near the Kraft break, evidenced by its significantly reduced spin-down rate (in the top two panels of that column).

\begin{figure}
      \centering
      \includegraphics[height=7.2cm]{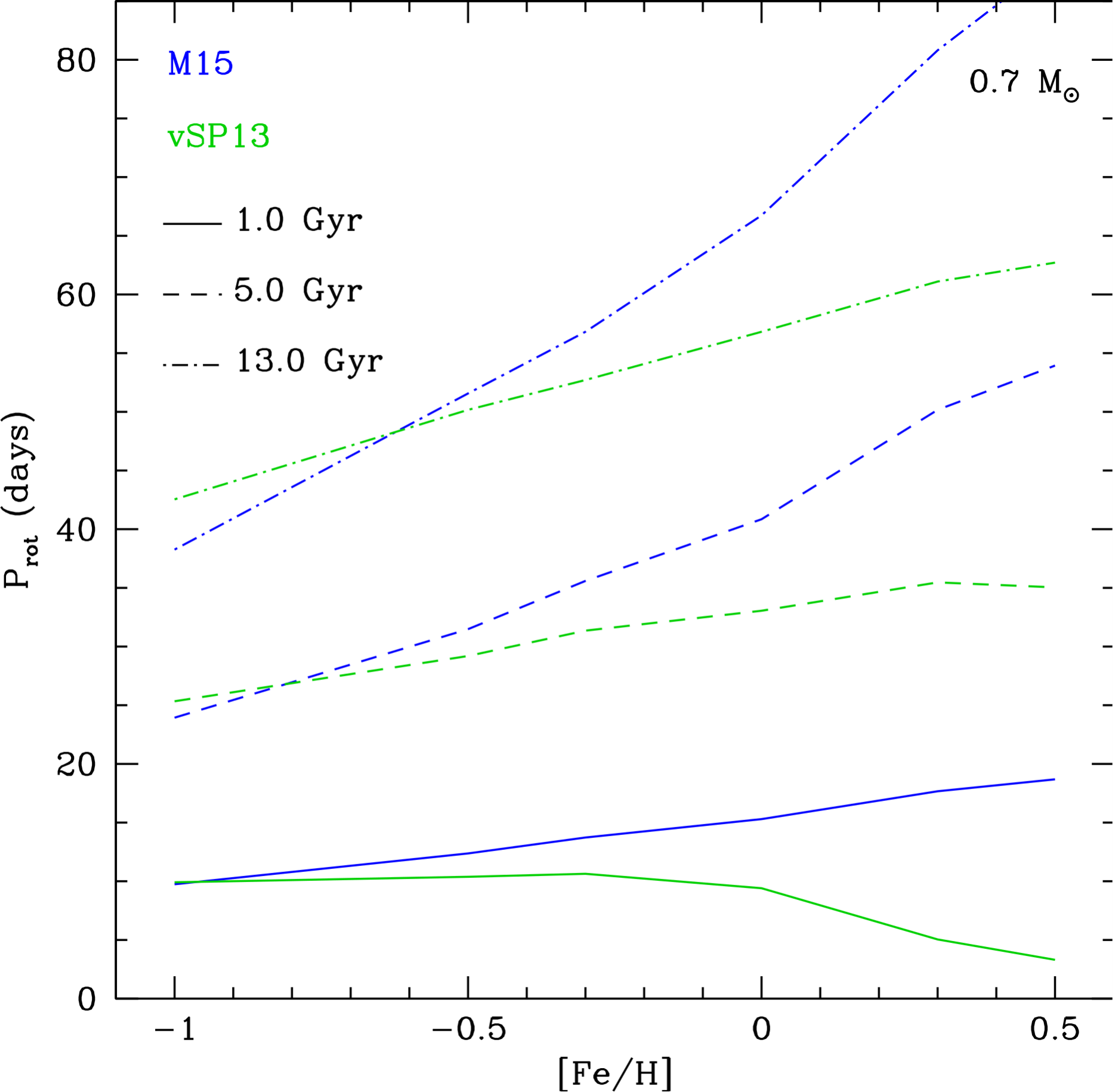}
      \includegraphics[height=7.2cm]{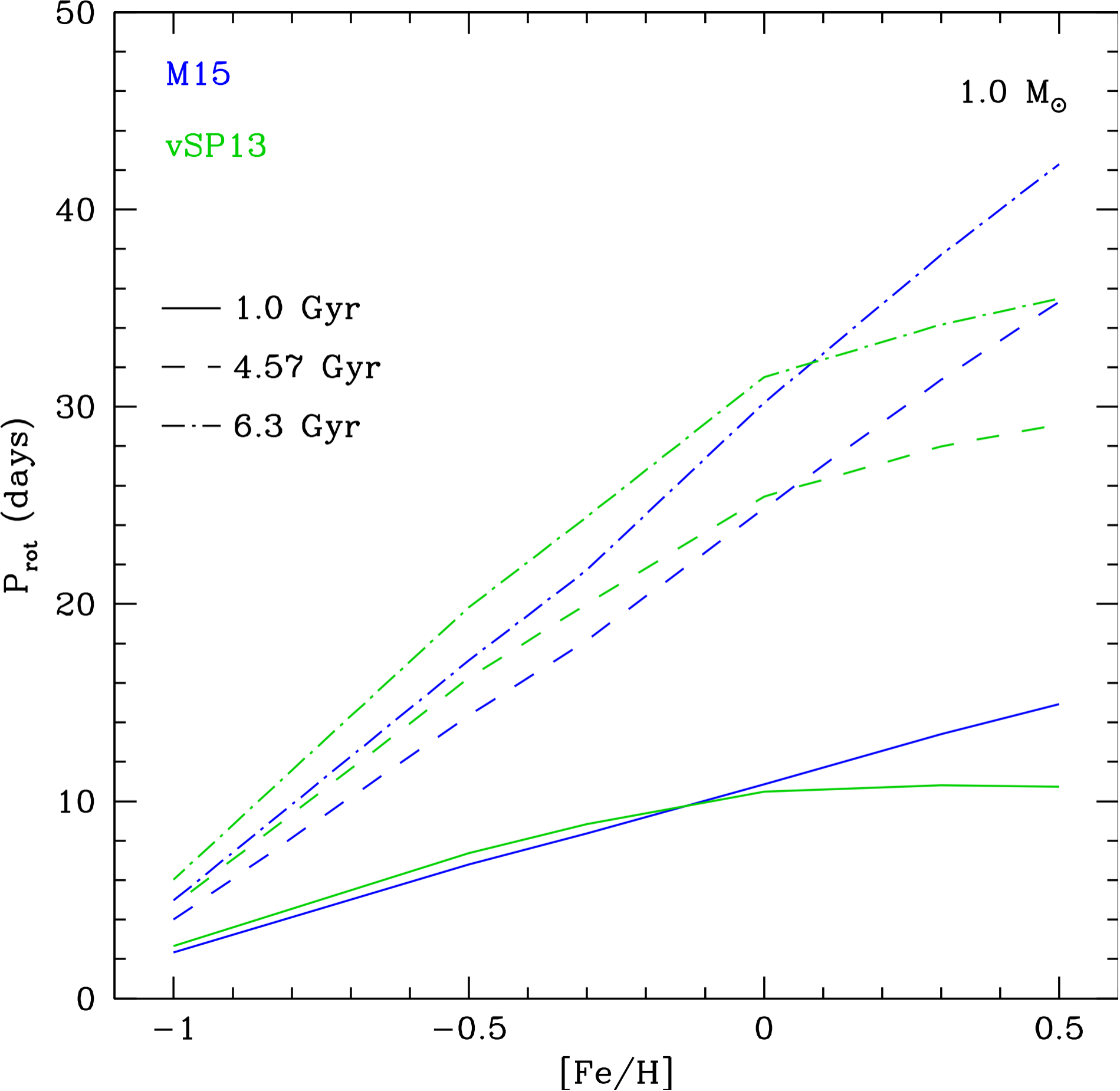}
      \includegraphics[height=7.2cm]{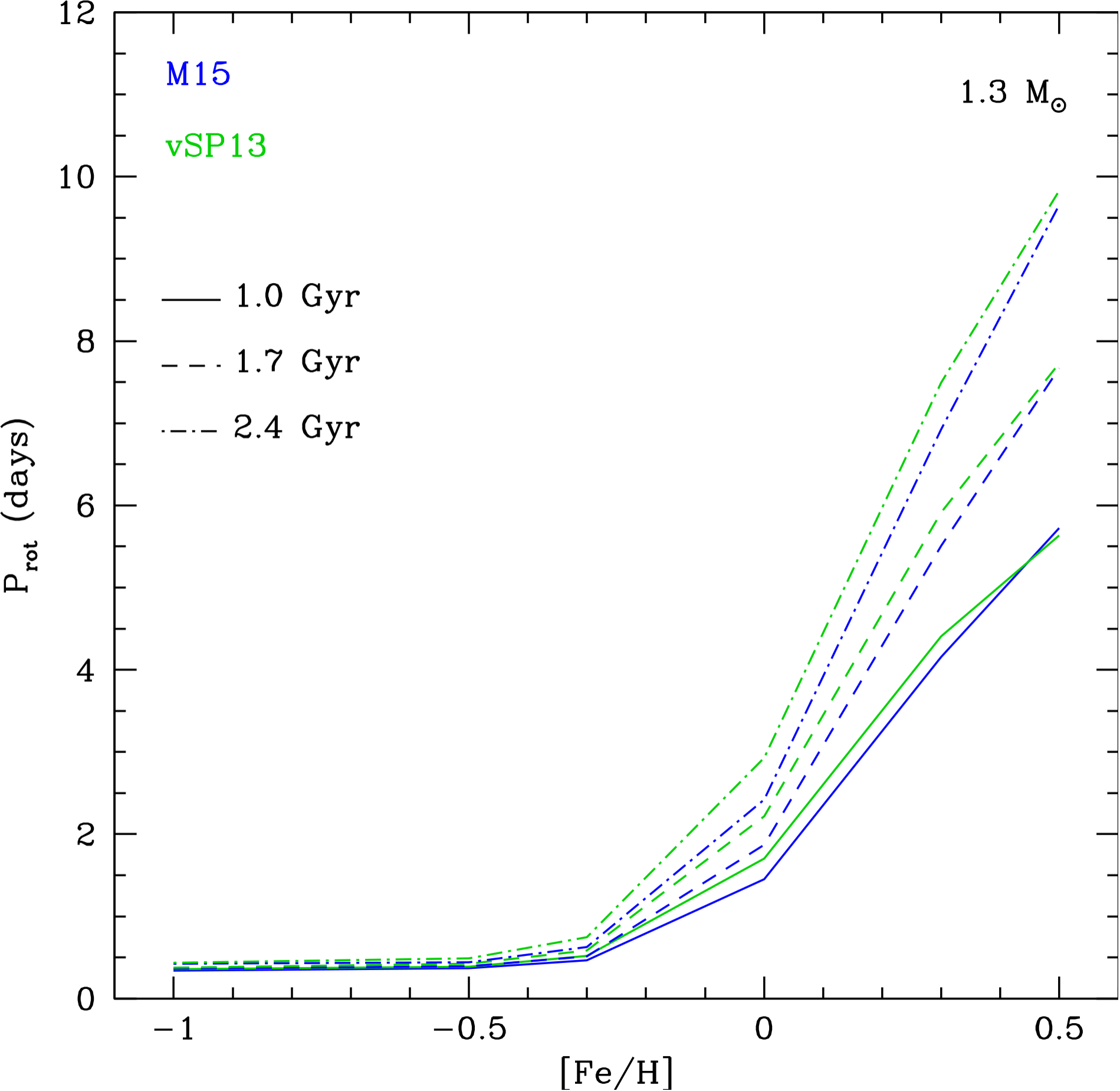}
      \caption{Predicted rotation period as a function of metallicity, shown at three different ages (as indicated) and computed using the torques of \cite{Mattetal2015} (blue lines) and \cite{VSP2013} (green lines).  The top, middle, and bottom panel show models for stars with 0.7M$_\odot$, 1.0M$_\odot$ and 1.3M$_\odot$, respectively.}
      \label{Fig:Prot_allmasses}
    \end{figure}

    Figure~\ref{Fig:Prot_allmasses} shows the predicted rotation period as a function of [Fe/H], at three fixed ages, for each of the three masses considered here, and for both the vSP (green lines) and M15 (blue lines) braking laws.  In addition to the trends described above, this space shows clearly the differences between the predictions of the the two braking laws.  Particularly for the 0.7M$_\odot$ case (top panel), the rotation periods obtained with the two models are very different at higher metallicity, with the M15 model predicting much slower rotators during the ages shown.  At 1.0~$M_\odot$ (middle panel), the differences are similar, although smaller in magnitude.  Finally, the 1.3M$_\odot$ models (bottom panel) predict very similar rotation periods for both braking laws, at all metallicities. Once again, we see that models with metallicities less than solar are not significantly spun down on the main sequence and a higher metallicity leads to a more important braking.

   \begin{figure}
      \centering
      \includegraphics[width=0.41\textwidth]{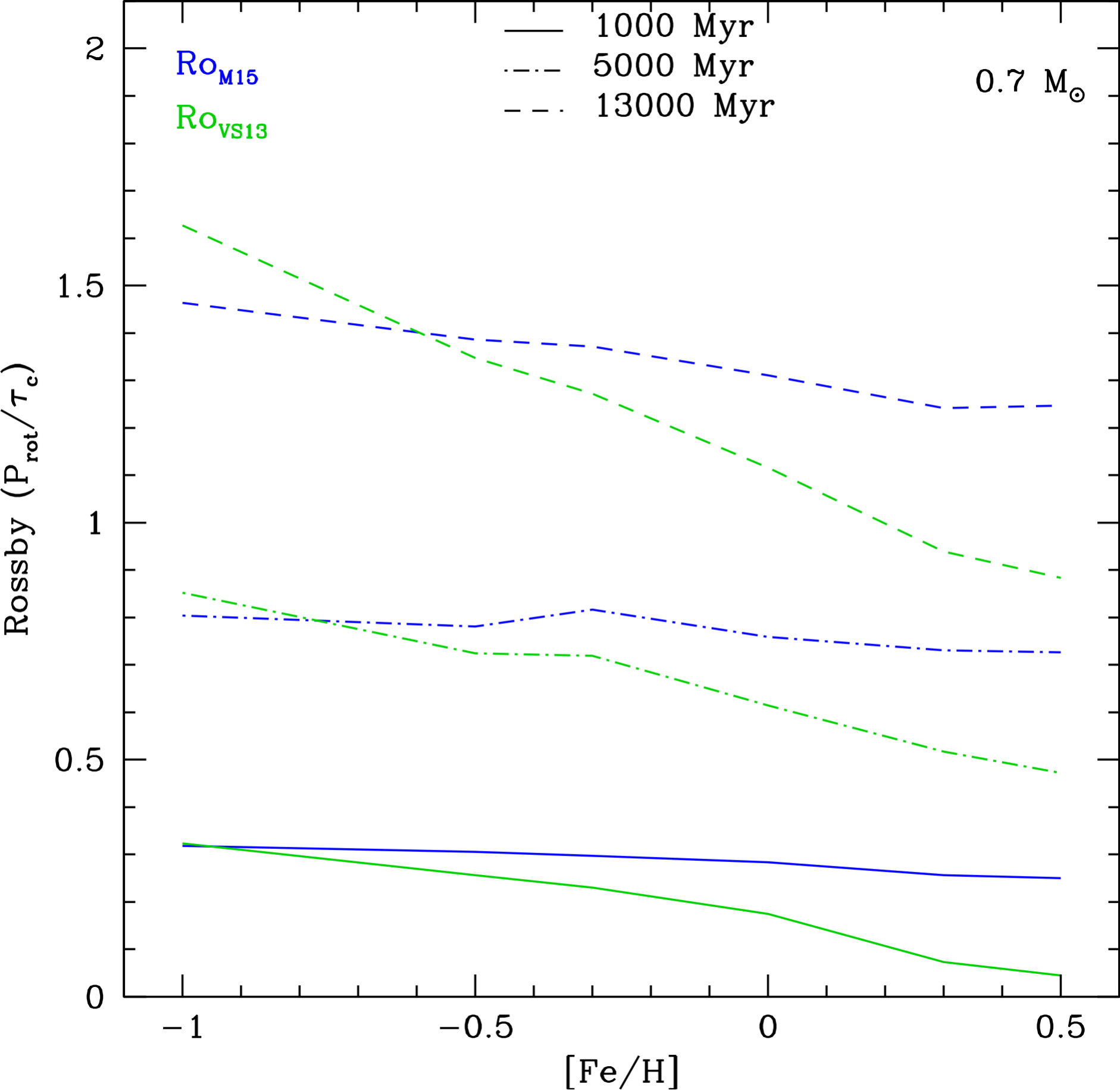}
      \includegraphics[width=0.41\textwidth]{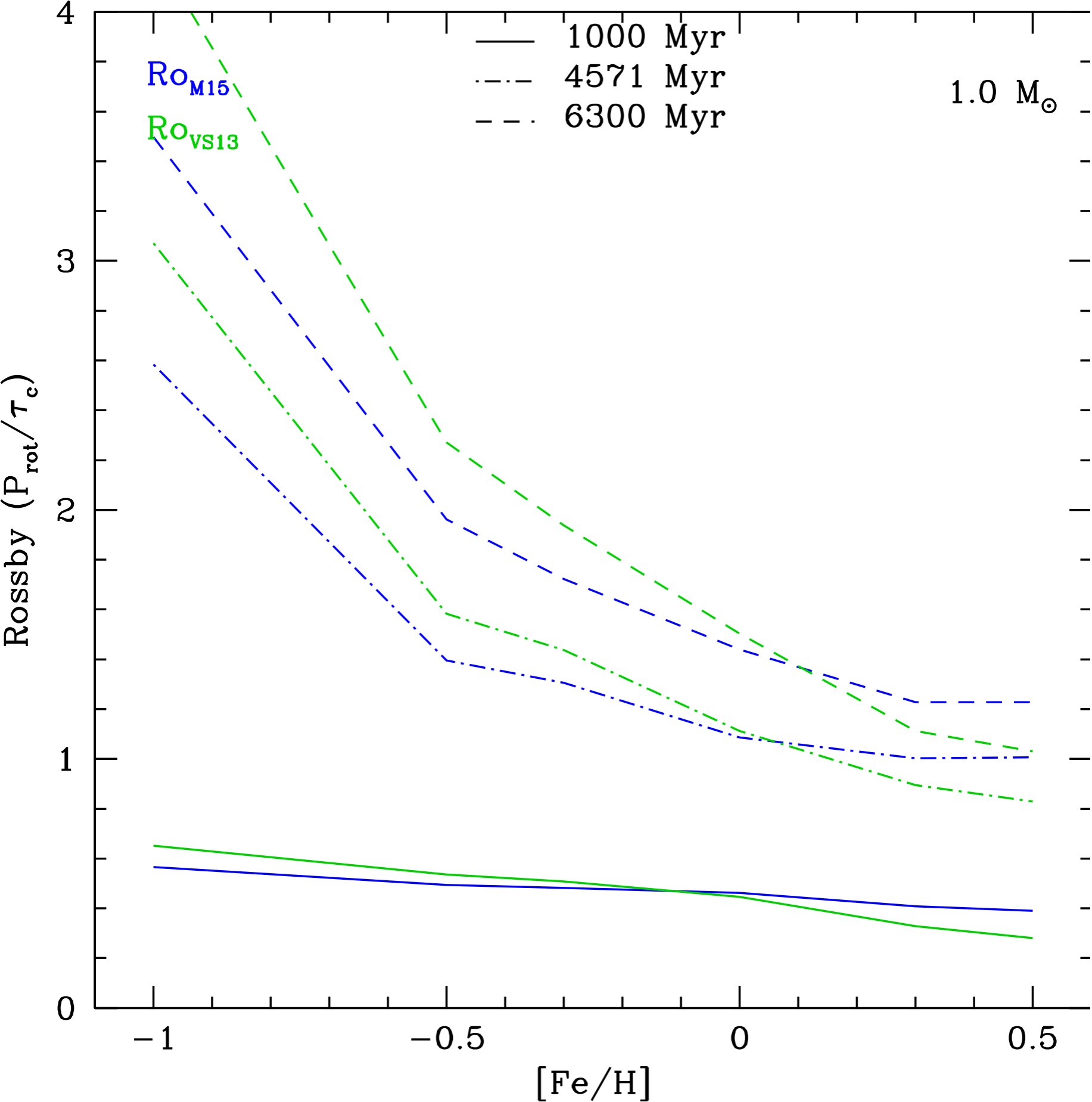}
      \includegraphics[width=0.41\textwidth]{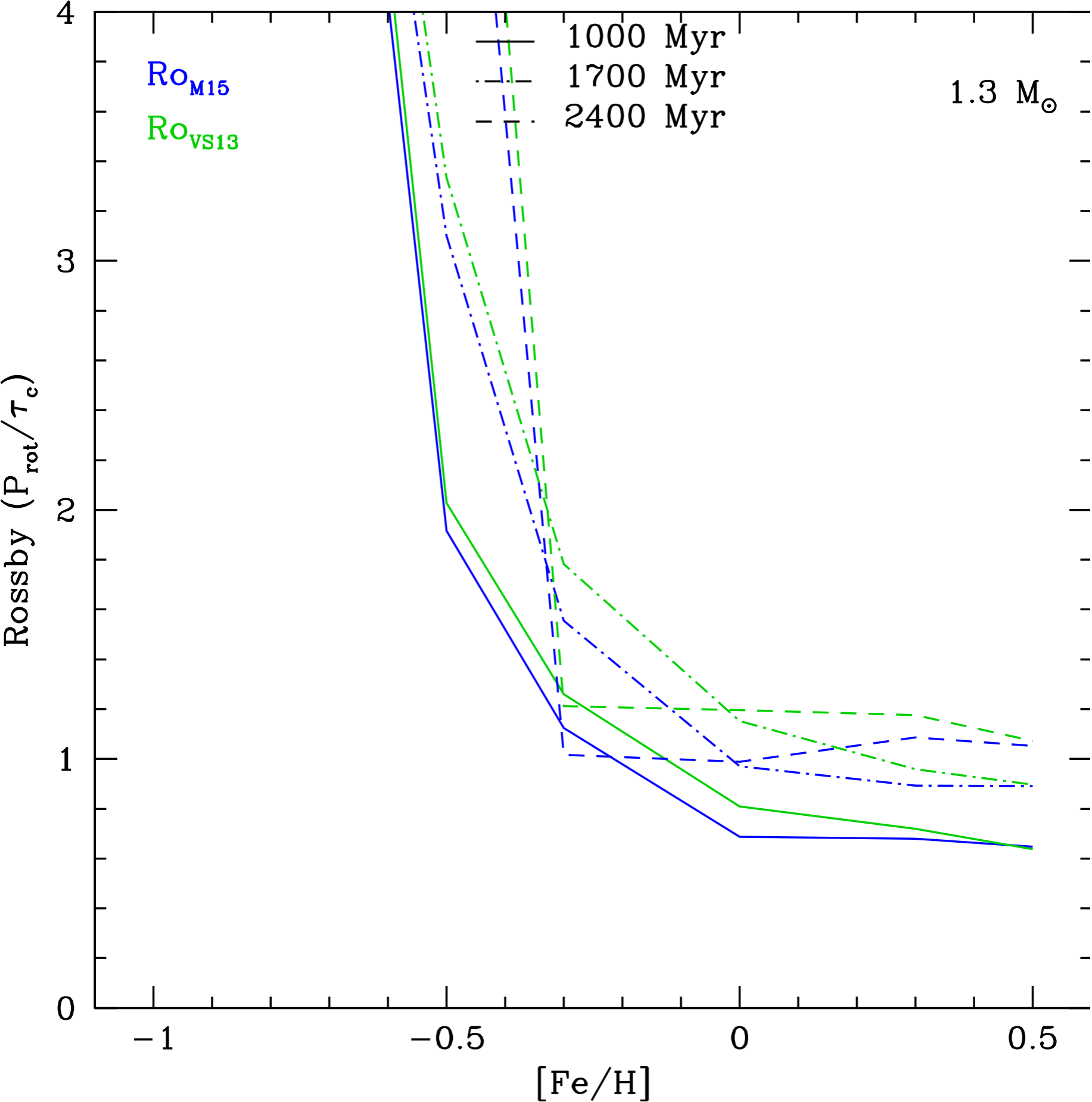}
      \caption{Predicted Rossby number as a function of metallicity, shown in the same format (and for the same models) as Figure \ref{Fig:Prot_allmasses}.  At a given age, stars with lower metallicity are predicted to have a larger Rossby number (and thus decreased magnetic activity), in spite of the fact that they are rotating more quickly (cf.\ Fig.\ \ref{Fig:Prot_allmasses}). 
      }
      \label{Fig:Rossby_allmasses}
    \end{figure}

  \subsection{Evolution of the Rossby number}
  \label{sect:rossby}

    As we saw from Fig.~\ref{Fig:param}, the main varying parameter affecting the torques is the convective turnover timescale. This is because, in both M15 and vSP13's torques, the important parameter is the Rossby number which is defined as the ratio between the convective turnover timescale and the rotation period. In Fig.~\ref{Fig:Rossby_allmasses} we present this parameter for both torques at a few selected ages along the main sequence for all masses and metallicities.  We saw on Fig.~\ref{Fig:Prot_allmasses} that the rotational evolution as well as the structural evolution is strongly affected by the metal-content of a star. 
    Both the rotation period and the convective turnover timescale increase with the metallicity of the star. Their ratio, the Rossby number is thus varying less than its two components, this can be seen on  Fig.~\ref{Fig:Rossby_allmasses}, in particular in the 0.7 M$_\odot$ diagram (top panel). 
     In this case, the Rossby number is almost constant with M+15's torque. For the 1M$_\odot$ model, despite a slower rotation rate, higher metallicity stars tend to have a smaller Rossby number due to their thicker convective envelope. The difference is even slightly increasing with age, from a factor 1.5 to 2.5 between the most metal-enriched to the most metal-poor models.  Finally for the more massive star (bottom), the increase of the rotation rate of the low-metallicity model doesn't compensate for the sharp drop-off of the convective turnover timescale as the convective envelope vanishes with metallicity. Thus no matter the metallicity, the 1.3 M$_\odot$ stars have a high Rossby number despite very low rotation periods.

    Thus, an interesting outcome of this work is that, due to the differences in the torques, the rotational evolution and trends with metallicity provided by the two torques is not the same, especially with lower mass-stars (see top panel in fig.~\ref{Fig:Prot_allmasses}). Consequently, the evolution of the Rossby number is expected to follow a different path with the two braking laws as displayed in fig~\ref{Fig:Rossby_allmasses}. 

\section{Discussion}
\label{sect:discussion}

  \subsection{Implications for Activity-Age Relationship}
  \label{sect:activityage}

    When looking at various indicators of magnetic activity such as X-rays \citep{Wright2011,Stelzer2016},  H$\alpha$ \citep{Newton2017} and H\&K index \citep{MamajekHillenbrand2008}, or stellar magnetic field properties \citep{See2015,See2019}, all correlate strongly with Rossby number \citep{Noyesetal84,Mittag2018}.  We showed in section \ref{sect:rossby} that the evolution of Rossby number depends strongly on metallicity, which implies that the decline of activity with age will happen very differently for stars with different metallicities.
    In particular, the prediction is that low metallicity stars will show a more rapid decline of activity with age, in spite of the fact that they do not spin down as quickly.
    Indeed, for a given mass, lower metallicity stars have a smaller convective turnover timescale, thus the ratio $P_\textrm{rot}/\tau_\textrm{cz} = Ro$ is changing faster with the rotation period for these stars.

    Activity level measurements might provide another way to test the models here, at least in overall trends in Rossby number. 
    At the same time, metallicity can affect other aspects of magnetic activity that we have not taken into account.  
    For example, \cite{Witzke2018} showed that, for a fixed activity level, metallicity affects the brightness contrasts between faculae and the photosphere, leading to differences in the optical variability and detectability of rotation.

    \cite{Karoff2018} studied a solar analog with the same age but twice the metallicity of the Sun ([Fe/H]=+0.3). 
    They found a magnetic cycle with a period around 7.5 years, slightly shorter than the one of the Sun, and with a higher amplitude of the brightness variability, indicating a higher level of magnetic activity.  
    The rotation period of the star is somewhat unclear.  
    \cite{Karoff2018} derived a strong differential rotation, with the equator and the pole completing a rotation in respectively 25 and 35 days and a mean rotation period of 27 days.
    \cite{Garciaetal2014} reported a period of $29.8\pm 3.1$ days.
    The rotation periods predicted by the vSP15 and M+15 torques (see Table~\ref{tab:Prot_sun}) are 27.5 and 31.4 days, respectively, which correspond to globally-averaged rates (because our models don't include differential rotation).  
    The uncertainties don't allow for a definite test of the predicted trends in rotation.
    However, as shown in section \ref{sect:rossby}, this star should have a smaller Rossby number than the Sun (see Fig.~\ref{Fig:Rossby_allmasses}, middle panel), in spite of it rotating at a similar rate. Thus, the activity level is expected to be higher, in qualitative agreement with the finding of \cite{Karoff2018}.
   
    \cite{Metcalfe2016} proposed that stellar spots vanish beyond a certain Rossby number around the solar value, and \cite{Vansaders2016} suggested that angular momentum loss becomes negligible beyond a similar value. Both were tested against the whole Kepler field in \cite{Vansaders2019}.
    If these ideas are correct, they may allow another way to distinguish between the two torque models.  The vSP13 torque predicts that metal rich stars are more active/faster rotators at older ages than the M+15 torque.  Thus the threshold beyond which stellar spots wouldn't appear anymore would be reached later in time. According to Figure~\ref{Fig:Rossby_allmasses}, for a 0.7M$_\odot$ at [Fe/H]=0.3, the Ro=1 threshold would be reached at respectively 4.5 and 8.5 Gyr for M15 and vSP13 prescriptions, which may be a measureable difference. 

  \subsection{Effect on planetary system and habitability}
    
    As describe in the previous section, our work suggests that metal rich stars will appear more active at a given rotation rate or age, having larger intrinsic variability. This is particularly relevant since planet detection is highly dependent on the stellar activity \citep{Queloz2001, MartinezArnaiz2011, Santerne2016} and that the effects of metallicity on planet formation and detection is gaining attention in the literature \citep[see \textit{e.g.}][]{Adibekyan2019, Sousa2019}.
    In particular, the detection probability of close-in massive planets has been shown to be directly proportional to the metallicity of the host-star \citep{Santos2001}. \cite{Cauley2018} on one side, and \cite{Strugarek2016} and \cite{Cauley2019} on the other side demonstrated recently that \LA{these types of planets may directly affect the rotation and the activity of the star through tidal interactions and magnetic reconnections, respectively.  Even though both mechanism are candidates, which mechanism dominates and how they actually work remains very unclear from a theoretical point of view.}
    Finally, if metal rich stars have enhanced magnetism, they may have a smaller habitable zone \cite{Vidotto2013, Johnstone2017, Gallet2017, Carolan2019}. 
    All these effects of metallicity on activity should be accounted for in the structural and rotational models that is used when selecting targets for planet characterization.

\subsection{Trends for Fixed $T_\textrm{eff}$}

\begin{figure}
        \centering
        \includegraphics[width=0.48\textwidth]{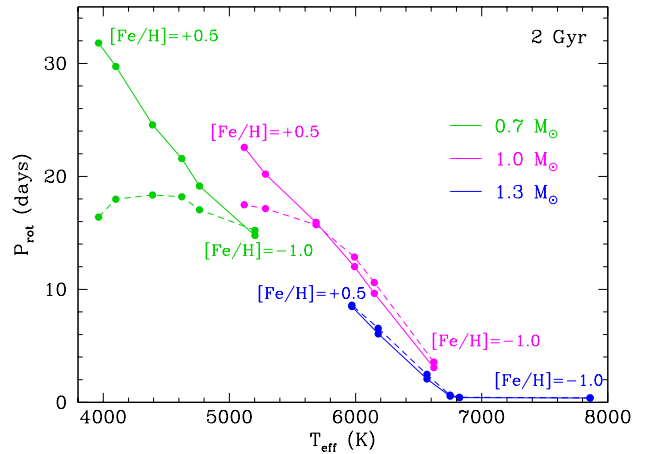}
        \caption{Rotation period as a function of the effective temperature at 2 Gyr for all the models computed using the torque of M15 (solid lines) and that of vSP13 (dashed lines).  Models with 0.7, 1.0 and 1.3M$_\odot$ are shown in green, magenta, and blue, respectively. For each mass, each point correspond to a different metallicity (high to low from left to right, as indicated).  Even at a fixed $T_\textrm{eff}$, metallicity affects the rotational evolution of stars, although the magnitude of the effect is smaller than for a fixed stellar mass.}
        \label{fig:Teffevol_Z}
    \end{figure}

  Metallicity strongly influences the effective temperature of a star. \cite{McQuillan2014} \eg suggested that effective temperature is a better tracer for rotation period than mass itself. 
  Indeed, it is a usual assumption to consider the convective turnover timescale as only a function of color or effective temperature \citep{Noyesetal84,CS2011}, and thus the torque and the rotational evolution depend mostly on T$_\textrm{eff}$. 
  On Fig~\ref{fig:Teffevol_Z} we plot the rotation period as a function of the effective temperature at 2 Gyr for the models of the three masses at all metallicity with the two modern torques.
  Indeed, a global trend is visible that shows cooler stars rotating slower on average. 
  That is, when considering all masses and metallicities, the dispersion in rotation period at a given effective temperature, is much lower than the spread in a P$_\textrm{rot}$-Mass diagram. 
  However, even for a given effective temperature, there is still a dependence on stellar mass (and associated [Fe/H]). 
  For example, according to the M15 torque, a 2Gyr old star with an effective temperature of 5200K can have a 14- or 21-day rotation period, depending whether its ($M_\star$, [Fe/H]) values are ($0.7$M$_\odot$, -1.0) or (1.0M$_\odot$, +0.4).
  In other words, the shape and location of gyrochrones  \citep[period-color or period-Teff relationships at fixed age; e.g.,][]{Barnes2010, Angus2015} will depend on metallicity.  
  As indicated in Figure~\ref{fig:Teffevol_Z}, the metallicity-dependence is complex (non-monotonic), so that gyrochrones at different metallicity can even cross one another. 

\subsection{Missing physics}

    The wind torques we have implemented were not derived with any direct dependencies on metallicity. 
    Thus, our models are likely missing some physical processes that are influenced by metallicity.
    First of all, we have assumed instantaneous internal redistribution of angular momentum, but physical transport processes can significantly influence the rotational evolution \citep{Denissenkov2010b, GB13, GB15, LS15, Amard2016, SL19}. 
    It is unclear how metallicity may affect internal transport, but it has a known strong influence (e.g.) on convection-zone depth, and thus likely is important. 
    Although it is unclear how metallicity may affect internal transport, \cite{Amard2019} showed that angular momentum transport by meridional circulation and shear-induced mixing is enhanced at lower metallicity, leading to a rotation profile closer to solid-body.  Additionally, the gradient of composition present in metal-poor stars is more important, and may prevent some hydro-dynamical processes to redistribute angular momentum in the radiative region of the star, already on the main sequence \citep{MeynetMaeder1997}.

    Another point that should be considered is the turnover timescale we've chosen. \cite{Hanasoge2012} showed that mixing-length theory provides good estimates of convective velocities and so of convective turnover timescale. 
    However, the location at which the convective velocity should be probed to characterise at best the dynamo properties remains a very uncertain parameter \citep[\textit{e.g.}][]{Landin2017,Charbonnel2017}. 
    We opted to use a half a pressure height scale above the convective bottom, but other choices may be equally valid.  
    No theoretical work has been done yet on the effect of stellar composition on dynamo and/or convection itself, so it is not clear what kind of influence it might have.

    Finally, the mass-loss rate is an important parameter, which is prescribed in different ways in the M15 and vSP torques. 
    The former assumes a Rossby-dependent (and possibly mass-dependent) $\dot{M}$, while the latter assumes it varies with the Rossby number and stellar luminosity \citep{Reimers1975}.
    \cite{Suzuki2018} modeled coronal winds of extremely metal-poor stars and found that, due to the reduced efficiency of radiative cooling in the corona, a lower metallicity leads to an enhanced mass-loss rate, which would increase the angular momentum loss and mitigate the effect we have seen in our models.

\section{Conclusions}
\label{sect:conclusion}

    In this work we combined the four main characteristics allowing to model a single star : its initial mass, chemical composition, rotation rate and its current age. 
    We have carried out for the first time a general theoretical study of the effect of six different metallicities on the rotational evolution of stars of three masses around a solar-mass. 
    We have shown that, according to recent angular momentum loss prescription implemented in spin evolution models, metallicity can have a strong impact on the rotational evolution of low-mass stars. 
    
    The classical torque model by \cite{Kawaler88} shows little variation with the chemical composition because it has no explicit dependence on the Rossby number.
    On the other hand, the torques by \cite{VSP2013} and \cite{Mattetal2015} present a strong variation with the Rossby number and thus convective turnover timescale\LA{, since these properties affect the dynamo action and the large-scale magnetic field. In particular,} the convective turnover timescale is a direct consequence of the stellar structure and thus is strongly affected by both the mass and the chemical composition of the star.
    
    We find that metal-poor stars are rotating faster than their solar metallicity counterpart of same mass and age but have a higher stellar Rossby number.
    In particular, stars with a higher mass than the Sun are very sensitive to metallicity because of their proximity to the Kraft break where the Rossby number is changing very quickly with the depth of the convective zone, and so with the chemical composition or the mass of the star. 
    
    Since stellar activity is somehow inversely proportional to the stellar Rossby number, our models predict the level of activity of fast rotating low-metallicity stars to be lower than stars with a higher metal-content at slower rotation. This prediction has already been partially confirmed by observations.

    Our results suggest that lower mass metal-rich stars would allow to discriminate between the torque descriptions by \cite{Mattetal2015} and \cite{VSP2013}.
    Finally, we would put a word of caution for the use of purely color-based gyrochrones since for a difference in metallicity of 0.3 dex, we found a possible 20\% error in terms of period for a given effective temperature at a fixed age.

    This work represents a necessary step toward a consistent and comprehensive modelling of the rotational evolution of low-mass star at metallicity other than solar. 
    It appears to be very timely since more complete data sets are appearing with large spectroscopic surveys such as APOGEE \citep{Majewski2017} and LAMOST \citep{Cui2012}, or the unWISE catalog \citep{Schlafly2019} to combined with GAIA DR2 parallaxes and rotation period \citep{Lanzafame2018}, Kepler, K2 and TESS rotation periods measurement for more field stars.

\acknowledgments
We would like to thank the referee for their very kind words. We acknowledge funding from the European Reseach Council (ERC) under the European Unions Horizon 2020 research and innovation program (grant agreement No 682393 AWESoMeStars). We would like to thank Aline Vidotto and the AWESoMeStars team member for useful discussions. This work also benefited from discussions within the international team “The Solar and Stellar Wind  Connection: Heating processes and angular momentum loss, supported by the International Space Science Institute (ISSI).

\bibliography{Biblio354}
\bibliographystyle{aasjournal}

\end{document}